\documentclass
[%
 aip,
 jmp,%
 amsmath,
 amssymb,
 preprint,%
 reprint,%
 onecolumn,%
]
{revtex4-1}
\usepackage{ulem}
\usepackage{graphicx}
\usepackage{dcolumn}
\usepackage{bm}
\usepackage[mathlines]{lineno}
\usepackage{color}%
\usepackage{caption}%
\usepackage{url}

\begin{document}

\title{A simple scheme of low phase noise microwave synthesizers based on the sub-sampling phase lock loop} %

\author{Wenbing Li}%
\thanks{Corresponding author E-mail: wenbing$\_$li@hust.edu.cn}
\affiliation{MOE Key Laboratory of Fundamental Physical Quantities Measurement, Hubei Key Laboratory of Gravitation and Quantum Physics, School of Physics, Huazhong University of Science and Technology, \\1037 Luoyu Road, Wuhan 430074, P. R. China}%

\author{Qiang Hao}%
\thanks{Corresponding author E-mail: haoq@ntsc.ac.cn}
\affiliation{Key Laboratory of Time and Frequency Primary Standards, National Time Service Center, Chinese Academy of Sciences, Xi'an 710600 P. R. China}%

\author{Yuanbo Du}%
\author{Shaoqing Huang}%
\affiliation{MOE Key Laboratory of Fundamental Physical Quantities Measurement, Hubei Key Laboratory of Gravitation and Quantum Physics, School of Physics, Huazhong University of Science and Technology, \\1037 Luoyu Road, Wuhan 430074, P. R. China}%

\author{Peter Yun}%
\affiliation{Key Laboratory of Time and Frequency Primary Standards, National Time Service Center, Chinese Academy of Sciences, Xi'an 710600 P. R. China}%

\author{Zehuang Lu}%
\thanks{Corresponding author E-mail: zehuanglu@hust.edu.cn}
\affiliation{MOE Key Laboratory of Fundamental Physical Quantities Measurement, Hubei Key Laboratory of Gravitation and Quantum Physics, School of Physics, Huazhong University of Science and Technology, \\1037 Luoyu Road, Wuhan 430074, P. R. China}%

\date{\today}%

\begin{abstract}

In this paper, we demonstrate a simple scheme of 6.835 GHz microwave frequency synthesizer based on the sub-sampling phase lock loop (PLL) technique. The application of the sub-sampling PLL is the key to simplify the architecture of the synthesizer in this scheme. A 100 MHz oven controlled crystal oscillator (OCXO) with ultra-low phase noise is used as the initial signal source. Then a dielectric resonant oscillator (DRO) of 6.8 GHz is directly phase locked to the 100 MHz OCXO utilizing the sub-sampling PLL. Benefiting from the sub-sampling PLL, the processes of microwave frequency multiplication and phase lock which are necessary in the development of microwave synthesizer are greatly simplified. Therefore, the architecture of the synthesizer is very simple. Correspondingly, the power consumption and cost of the synthesizer are low. The absolute phase noises of the 6.835 GHz output signal are measured to be -47 dBc/Hz, -77 dBc/Hz, -104 dBc/Hz and -121 dBc/Hz at 1 Hz, 10 Hz, 100 Hz and 1 kHz offset frequencies, respectively. he synthesizer can be used as the local oscillator of the Rb atomic clocks. For the Rb atomic clocks operated in the continuous or pulsed optically pumped (POP) mode, Tthe short-term frequency stability limited by the absolute phase noises of the synthesizer through the intermodulation or the Dick effect is theoretically calculated to be better than $5.0 \times 10^{-14}\tau^{1/2}$. This low phase noise microwave frequency synthesizer can be used in other experiments of fundamental physics measurement.

\end{abstract}
\maketitle

\section{INTRODUCTION}

Low phase noise microwave frequency synthesizers are generally used in the experiments of Raman atomic interferometer,\cite{Hu2013} testing local position invariance,\cite{Ashby2007} precision atomic spectroscopy measurement,\cite{Fortier2007} atomic spin squeezing,\cite{Chen2012} and atomic clock.\cite{Du2015,Levi2014,Fernando2010,Francois2014,Boudot2009,Li2018,Francois2015} The phase noise of the microwave synthesizer is one of the key factors in limiting the precision of those measurements.

Among the atomic clocks, the Rb atomic clocks are widely used in the fields of telecommunications, global navigation satellite system (GNSS) and industrial applications because of its superior characteristic of high frequency stability, low power consumption, and small volume.\cite{Camparo2007,Vannicola2010,Micalizio2015} In recent years with widespread applications promoted by GNSS, the frequency stability of the Rb atomic clocks has been greatly improved. For example, the short-term frequency stability of the Rb atomic clocks operated in a continuous optically pump mode has been better than $ 2.5\times10^{-13}\tau^{-1/2}$.\cite{Bandi2014,Hao2016} A pulsed optically pumped (POP) Rb atomic clock with short-term frequency stability of $ 1.7\times10^{-13}\tau^{-1/2}$ has been developed.\cite{Calosso2012}

However, no matter whether the Rb atomic clocks operated in the mode of continuous optically pumping or POP, the absolute phase noise of local oscillator is still one of the main factors to limit their short-term frequency stability. This has been analyzed and verified by several researches.{\cite{Dick1987,Deng1999,Joyet2001}} For the continuous interrogation atomic clocks, the frequency modulation noise of the local oscillator at even harmonics offset frequencies would affect its short-term frequency stability via the intermodulation effect.{\cite{Deng1999}} For the POP clocks, its short-term frequency stability would be affected by the absolute phase noise of the local oscillator through the Dick effect.{\cite{Dick1987}} For both type Rb atomic clocks, their short-term frequency stabilities would be limited by the absolute phase noise of the local oscillator, especially at the offset frequency range of 100 Hz to 1 kHz. Therefore, in order to improve the short-term frequency stability of Rb atomic clocks, it is beneficial to develop a synthesizer with low phase noise at these offset frequencies.

TThe short-term frequency stability of the atomic clocks could be considerably improved by the ultra-low phase noise of the ultra-stable laser {\cite{millo2009,Fortier2011}} or cryogenic sapphire oscillator (CSO) based synthesizer.{\cite{Lipphardt2008,Abgrall2016,Takamizawa2014}} Nevertheless, for the compact Rb clocks used in the GNSS or industry applications, it is not realistic to use these complex, expensive and bulky systems. Thus, it will still remain a well-adapted solution to develop the ultra-low noise quartz crystal oscillator based microwave frequency synthesizer, and there have been several reports on developing microwave synthesizers for atomic clocks.\cite{Fernando2010,Francois2014,Boudot2009,Li2018,Francois2015,Heavner2005} In most of these schemes, to phase lock a dielectric resonant oscillator (DRO), the 100 MHz OCXO is usually frequency multiplied to 6.8 GHz (for the Rb atomic clocks) or 9.2 GHz (for the Cs atomic clocks) with a non-linear transmission line (NLTL) or a Step Recovery Diode (SRD), and then mixed with the output signal of the DRO to generate an intermediate frequency signal which is further phase locked to a DDS, whose referenced signal is also provided by the OCXO. In that way the output frequency of the DRO is indirectly phase locked to the 100 MHz of OCXO. In this process, our previous work shows that the additional phase noise of the NLTL are greatly affected by its operation parameters such as the input power, the input and output impedances.{\cite{Li2018} To suppress the additional phase noise of the NLTL, the architecture of the schemes would be very complex, bulky, and expensive for optimizing the operation parameters of the NLTL. To simplify the structure, B. Fran\c cois \textit{et al.} designed a simple microwave synthesizer with a 1.6 GHz custom-designed frequency multiplication module. But the subsequent frequency division and multiplication are still needed.} Therefore, the structure of the synthesizer is still quite complicated.\cite{Francois2015} In this paper, we demonstrate a simple scheme of a 6.835 GHz low phase noise microwave synthesizer based on the sub-sampling PLL for Rb atomic clock. With the help of the sub-sampling PLL technique, which is usually used in the radio-frequency and radar engineering\cite{Gao2015,Gao2009}, the structure of the synthesizer can be greatly simplified.T And the absolute phase noise is a litte better than the result in ref[8]. The microwave frequency  synthesizer developed here can also be used in the experiments of the other experiments of fundamental physics measurement. 

\section{Scheme of the synthesizer}

The architecture of the synthesizer is shown in Fig. 1. The 100 MHz OCXO (Beijing Institute of Radio Metrology and Measurement, ZF550) is divided into two arms via a power splitter (Mini-Circuits, ZX10-2-12+). The first arm 100 MHz signal is used as the reference signal for a home-made direct digital synthesizer (DDS) that is based on AD9852 chip, which is used to achieve functions of frequency modulation and scanning. The DDS is controlled through serial communication with a computer. The second arm 100 MHz signal is used to phase lock the 6.8 GHz DRO utilizing a sub-sampling PLL. With the help of the sub-sampling PLL technique, the DRO is directly phase locked to the 100 MHz OCXO. The microwave frequency multiplication and mixing process are greatly simplified. The phase locked DRO is labelled as PLDRO. The 6.8 GHz output signal of the PLDRO is passed through a microwave isolator (Fairview, FMIR1021) to prevent the microwave signal from being reflected back into the PLDRO. The output signal of the PLDRO is then mixed with the 34.68 MHz output signal of the DDS with a mixer (Mini-Circuits, ZX05-83+) to generate the final 6.835 GHz output signal. Compared to the previous works on the development of microwave synthesizer, the structure of the synthesizer of Fig. 1 is much simpler. Correspondingly, the power consumption and the cost of the synthesizer are considerably decreased.\cite{Fernando2010,Francois2014,Francois2015,Li2018}

The detailed schematic of the sub-sampling PLL is shown in Fig. 2. The sampling phase detector is the key component of the PLL. The sampling phase detector mainly consisted of a step recovery diode (SRD) and a pair of Schottky diodes. The maximum frequency of the phase detector can reach 18 GHz. The input impedance of the SRD is usually about 4 - 10 $\Omega$, but the output impedance of the 100 MHz OCXO is usually 50 $\Omega$. Thus, we design an input matching circuit to match their impedances, which is realized by a transformer. In addition for matching impedance, the transformer is also used to convert unipolar pulses into symmetrical pulses. The loop pass filter (LPF) of the sub-sampling PLL is realized by a positive proportional-integral (PI) circuit. A voltage follower behind the sampling phase detector is designed to serve functions of isolation and impedance matching between the sampling phase detector and the LPF, and also as a low pass filter to suppress the high-frequency noise of the PLL. The scanning circuit is designed to improve the capture range of the PLL and to shorten phase locking time. The volume of the developed PLDRO is only slightly larger than that of a microwave amplifier. Thus, the volume of the synthesizer is reduced.

\begin{figure}
\includegraphics[width=14cm]{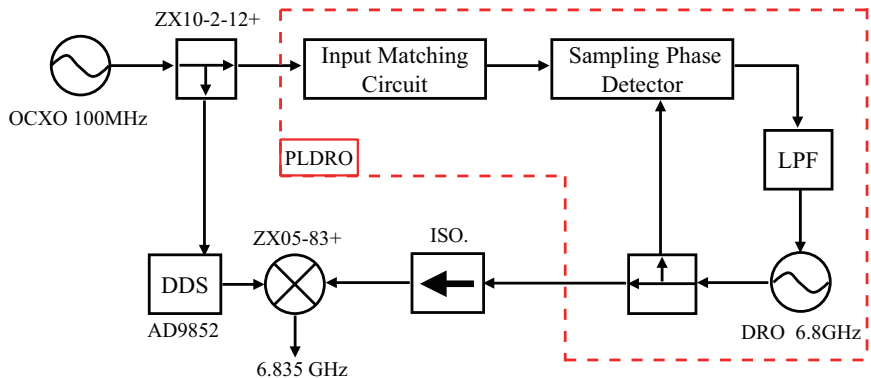}
\caption{(Colors online) The architecture of the microwave synthesizer, the part enclosed in the red dashed line is the phase locked DRO, labelled as PLDRO.}
\end{figure}

\begin{figure}
\includegraphics[width=14cm]{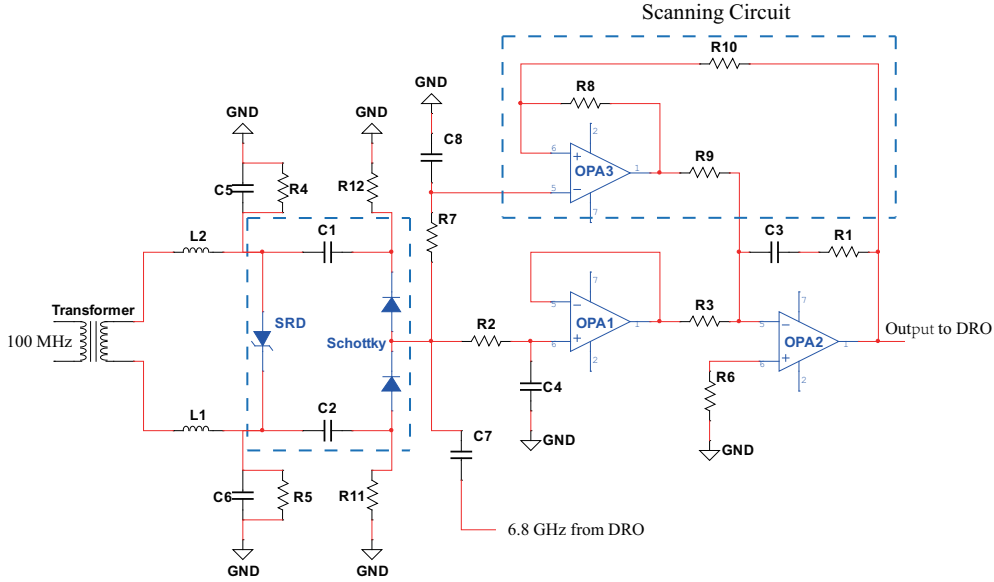}
\caption{(Colors online) The detailed schematic of the sub-sampling PLL}
\end{figure}

\begin{table*}
\renewcommand{\arraystretch}{1.5}
\caption{The phase noise sensitivity floor of the FSWP26 with option B61 (Xcorr=1, dBc/Hz)}
\centering
\begin{tabular}{ccccccccc}
  \hline
\textbf{Carrier frequency}&\textbf{1 Hz}&\textbf{10 Hz}&\textbf{100 Hz}&
\textbf{1 kHz}&\textbf{10 kHz}&\textbf{100 kHz}&\textbf{1 MHz}\\ \hline
100 MHz&-92&-115& -140 & -166&-170&-173&-175\\
7 GHz&-55&-78&-103&-133&-152&-153&-157\\ \hline
\end{tabular}
\end{table*}

\section{Results and discussion}

We measure the absolute phase noises of the main output signals of the microwave synthesizer at carrier frequencies of 100 MHz, 35 MHz, 6.8 GHz and 6.835 GHz with a Phase Noise Analyzer (R$\&$S FSWP26 with option 61) using cross-correlation technique for enhanced sensitivity. The noise floor of the FSWP26 with option B61 is shown in Table I (taken from FSWP26 datasheet). The numbers given in Table I is given under the condition of the cross-correlation factor of one (Xcorr=1). The noise floor can be further lowered with an increasing Xcorr as 5*log(Xcorr). In the actual measurement, the Xcorr and the bandwidth are set  as 50 and 3$\%$, respectively. The phase noise floor are -55 dBc/Hz, -85 dBc/Hz, -114 dBc/Hz, -138 dBc/Hz and -142 dBc/Hz at 1 Hz, 10 Hz, 100 Hz, 1 kHz and 10 kHz offset frequencies, respectively, which will not limit the measurement accuracy of the phase noise. The measured phase noises are shown in Fig. 3. The absolute phase noise of the OCXO 100 MHz has a slight bump at the offset frequencies around 100 kHz, which is inherent in the custom-made free running OCXO. The phase noise of the 6.8 GHz signal of the PLDRO at offset frequencies smaller than 100 Hz is deteriorated by an idealized 20*log(N). At offset frequencies larger than 100 Hz, especially at the offset frequency of locking bandwidth, the maximum phase noise deterioration is increased by an additional 5 dB. This is because the absolute phase noise of free running DRO at small offset frequencies is quite poor. The locking bandwidth of PLL is about 150 kHz. Thus, the gain within the servo loop bandwidth is enough. As a consequence, the phase noise, especially at the offset frequencies near the locking bandwidth, cannot be sufficiently suppressed. Even with this limitation, the phase noise performances are still better than the results of the scheme based on the NLTL or SRD. That is because the impedance matching is much better than the NLTL based scheme. The phase noise of the 6.835 GHz output signal is the same as the phase noise of the PLDRO. The phase noise of the DDS 35 MHz signal is excellent, which does not limit the phase noise of the  6.835 GHz output signal. Using the absolute phase noise of the 6.835 GHz output signal. the short-term frequency stability of the atomic clock is estimated to be better than $5.0 \times 10^{-14}\tau^{1/2}$.

\begin{figure}
\includegraphics[width=14cm]{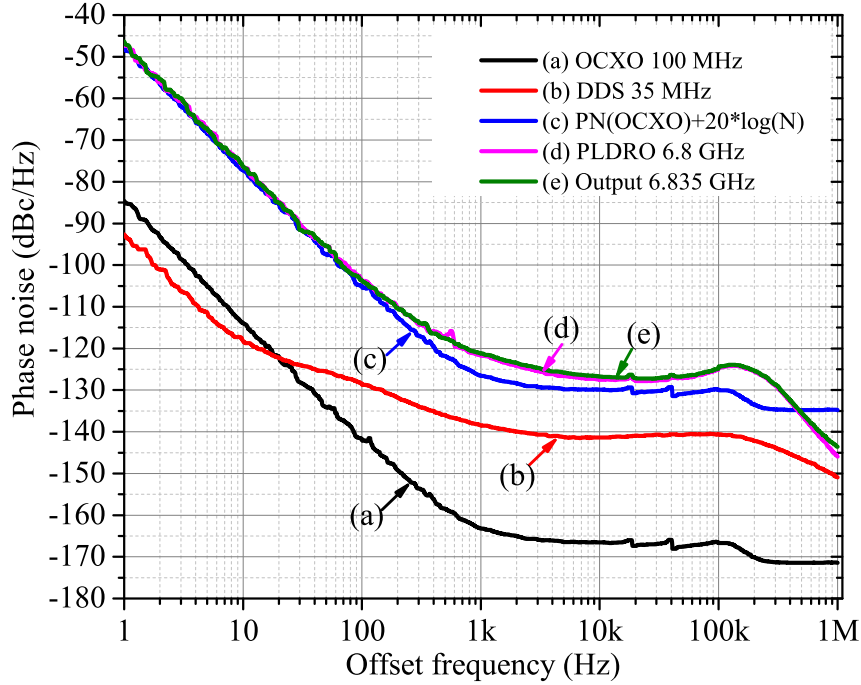}
\caption{(Colors online) The absolute phase noises of the main output signals from the microwave synthesizer. The carrier frequencies are at: (a) 100 MHz of the OCXO, (b) 35 MHz of the DDS, (c) 6.8 GHz deteriorated by idealized 20*log(N) from 100 MHz OCXO, }(d) 6.8 GHz of the PLDRO, (e) 6.835 GHz output signal.
\end{figure}

In conclusion, we demonstrate a simple low phase noise 6.835 GHz microwave synthesize based on the technique of sub-sampling PLL, with the associated low power consumption and cost. The absolute phase noises of the 6.835 GHz output signal are measured to be -47 dBc/Hz, -77 dBc/Hz, -104 dBc/Hz and -121 dBc/Hz at 1 Hz, 10 Hz, 100 Hz and 1 kHz offset frequencies, respectively. When it is used as the local oscillator of the Rb atomic clocks, the short-term frequency stability limit of the Rb atomic clocks due to the absolute phase noise via the intermodulation or the Dick effect is theoretically estimated to be better than $5.0 \times 10^{-14}\tau^{1/2}$. While in this work we only develop a microwave synthesizer for Rb atomic clocks, the same scheme can be easily adapted to be used for Cs atomic clocks by simply changing the output frequencies of the PLDRO and DDS. The microwave frequency  synthesizer can also be used in the experiments of the Raman atomic interferometers or gyroscopes and other experiments of fundamental physics measurement.  Limited by the phase noise of the DRO and gain and bandwidth of the PLL, the phase noise performance of the synthesizer realized via this method would not expect to reach the performance of the ultra-stable laser or CSO based synthesizer.

This work was partially supported by the National Key R$\&$D Program of China (Grant No. 2017YFA0304400), the National Natural Science Foundation of China (Grant No. 91336213, 11703031, U1731132 and 11774108) and the Post-doctoral Science Foundation of China (Grant No. 2017M612431).


\begin{thebibliography}{15} 

\bibitem[1]{Hu2013}
Z. K. Hu, B. L. Sun, X. C. Duan, M. K. Zhou, L. L. Chen, S. Zhan,Q. Z. Zhang and J. Luo, Phys. Rev. A \textbf{88}, 043601 (2013).
%
\bibitem[2]{Ashby2007}
N. Ashby, T. P. Heavner, S. R. Jefferts, T. E. Parker, A. G. Radnaev, and Y. O. Dudin, Phys. Rev. Lett. \textbf{98}, 070802(2007)
%
\bibitem[3]{Fortier2007}
T. M. Fortier, N. Ashby, J. C. Bergquist, M. J. Delaney, S. A. Diddams, T. P. Heavner, L. Hollberg, W. M. Itano, S. R. Jefferts, K. Kim, F. Levi, L. Lorini, W. H. Oskay, T. E. Parker, J. Shirley, and J. E. Stalnaker, Phys. Rev. Lett. \textbf{98}, 070801(2007)
%
\bibitem[4]{Chen2012}
Z. L. Chen, J. G. Bohnet, J. M. Weiner and J. K. Thompson, Rev. Sci. Instrum. \textbf{83}, 044701 (2012).
%
\bibitem[5]{Du2015}
Y. B. Du, R. Wei, R. C. Dong, F. Zou, and Y. Z. Wang, Chin. Phys. B \textbf{24}, 070601 (2015).
%
\bibitem[6]{Levi2014}
F. Levi, D. Calonico, C. E. Calosso, A. Godone, S. Micalizio and G. A. Costanzo, Metrologia \textbf{51}, 270(2014).
%
\bibitem[7]{Fernando2010}
F. Ram\'irez-Martinez, M. Lours, P. Rosenbusch, F. Reinhard and J. Reichel, IEEE Trans. Ultrason.  Ferroelectr. Freq. Control \textbf{57}, 88(2010).
%
\bibitem[8]{Francois2014}
B. Fran\c cois, C. E. Calosso, J. M. Danet and R. Boudot, Rev. Sci. Instrum. \textbf{85}, 094709 (2014).
%
\bibitem[9]{Boudot2009}
R. Boudot, S. Guerandel and E. De Clercq, IEEE Trans. Instrum. Meas. \textbf{58}, 3659 (2009).
%
\bibitem[10]{Li2018}
W. B. Li, Y. B. Du, H. Li and Z. H. Lu, AIP Adv. \textbf{8}, 095311 (2018).
%
\bibitem[11]{Francois2015}
B. Fran\c cois, C. E. Calosso, M. Abdel Hafiz, S. Micalizio and R. Boudot, Rev. Sci. Instrum. \textbf{86}, 094707 (2015).

\bibitem[12]{Heavner2005}
T. P. Heavner, S. R. Jefferts, E. A. Donley, T. E. Parker, F. Levi, Proceedings of the 2005 IEEE International Frequency Control Symposium and Exposition, 308(2005)
\textbf{86}, 094707 (2015).
%
\bibitem[12]{Camparo2007}
J. C. Camparo, Phys. Today \textbf{60}, 33 (2007).
%
\bibitem[13]{Vannicola2010}
F. Vannicola, R. Beard, J. White, K. Senior, M. Largay and J. A. Buisson, in proceedings of 42nd PTTI System and Applications Meeting, 181 (2014).
%
\bibitem[14]{Micalizio2015}
S. Micalizio, F. Levi, A. Godone, C. E. Calosso and I. Nazionale, in proceedings of IFCS and EFTF, 1 (2015).
%
\bibitem[15]{Bandi2014}
T. Bandi, C. Affolderbach, C. Stefanucci, F. Merli, A. K. Skrivervik and G. Mileti, IEEE Trans. Ultrason. Ferroelectr. Freq. Control \textbf{61}, 1769 (2014).
%
\bibitem[16]{Hao2016}
Q. Hao, W. Li, S. He, J. Lv, P. Wang and G. Mei, Rev. Sci. Instrum. \textbf{87}, 123111 (2016).
%
\bibitem[17]{Calosso2012}
C. E. Calosso, A. Godone, F. Levi and S. Micalizio, IEEE Trans. Ultrason. Ferroelectr. Freq. Control \textbf{59}, 2646 (2012).
%
\bibitem[18]{Deng1999}
J. Q. Deng, G. Mileti, R. E. Drullinger, D. A. Jennings and F. L. Walls, Phys. Rev. A \textbf{59}, 773 (1999).
%
\bibitem[19]{Dick1987}
G. J. Dick, in Proceedings of PTTI \textbf{19}, 133 (1987).
%
\bibitem[20]{Joyet2001}
A. Joyet, G. Mileti, G. Dudle and P. Thomann, IEEE Trans. Instrum. Meas.\textbf{50}, 150 (2001).
%
\bibitem[21]{millo2009}
J. Mill\"o, R. Boudot, M. Lours, P. Y. Bourgeois, A. N. Luiten, Y. L. Coq, Y. Kersal\'e and G. Santarelli, Opt. Lett. \textbf{34}, 3707 (2009).

\bibitem[22]{Fortier2011}
T. M. Fortier, M. S. Kirchner, F. Quinlan, J. Taylor, J. C. Bergquist, T. Rosenband, N. Lemke, A. Ludlow, Y. Jiang and C. W. Oates, Nature Photonics \textbf{5}, 425 (2011).

\bibitem[23]{Lipphardt2008}
B. Lipphardt, G. Grosche, U. Sterr, C. Tamm, S. Weyers and H. Schnatz, IEEE Trans. Instru. Meas.  \textbf{58}, 1258 (2008).
%
\bibitem[24]{Abgrall2016}
M. Abgrall, J. Gu\'ena, M. Lours, G. Santarelli, M. E. Tobar, S. Bize, S. Grop, B. Dubois, C. Fluhr and V. Giordano, IEEE Trans. Ultrason. Ferroelectr. Freq. Control \textbf{63}, 1198 (2016).
%
\bibitem[25]{Takamizawa2014}
A. Takamizawa, S. Yanagimachi, T. Tanabe, K. Hagimoto, I. Hirano, K. Watabe, T. Ikegami and J. G. Hartnett, IEEE Trans. Ultrason. Ferroelectr. Freq. Control \textbf{61}, 1463 (2014).

\bibitem[26]{Gao2015}
X. Gao, E. Klumperink and B. Nauta,  2015 IEEE Custom Integrated Circuits Conference (CICC), (2015).

\bibitem[27]{Gao2009}
X. Gao, G. Eric A. M. Kpluperink, Paul F. J. Geraedts and B. Nauta, IEEE Trans. Circuits and Systems II: Express Briefs  \textbf{56}, 117 (2009).

\end{thebibliography}
\end{document}